\documentclass[aps,pre,twocolumn,showpacs,floatfix,superscriptaddress]{revtex4-1}
\usepackage{graphicx}
\usepackage{subfigure}
\usepackage{url}
\usepackage{hyperref}

\usepackage{inconsolata}
\usepackage{amsmath}

\usepackage{braket}
\usepackage[capitalize]{cleveref}
\crefname{section}{Sec.}{Secs.} 

\begin{document}

\title{Metamagnetism and zero-scale-factor universality in the two-dimensional $J$-$Q$ model}


\author{Adam Iaizzi}
\email{iaizzi@bu.edu}
\homepage{www.iaizzi.me}
\affiliation{Department of Physics, Boston University, 590 Commonwealth Avenue, Boston, Massachusetts 02215, USA}
\affiliation{Center of Theoretical Sciences and Department of Physics, National Taiwan University, No. 1, Section 4, Roosevelt Road, Taipei 10607, Taiwan}

\author{Kedar Damle}
\affiliation{Department of Theoretical Physics, Tata Institute of Fundamental Research, Mumbai 400 005, India}

\author{Anders W Sandvik}
\affiliation{Department of Physics, Boston University, 590 Commonwealth Avenue, Boston, Massachusetts 02215, USA}
\affiliation{Beijing National Laboratory for Condensed Matter Physics and Institute of Physics, Chinese Academy of Sciences, Beijing 100190, China}


\date{\today}
\begin{abstract}
Using a combination of quantum Monte Carlo and exact methods, we study the field-driven saturation transition of the two-dimensional $J$-$Q$ model, in which the antiferromagnetic Heisenberg exchange $(J)$ coupling competes with an additional four-spin interaction $(Q)$ that favors valence-bond solid order. 
For small values of $Q$, the saturation transition is continuous, and is expected to be governed by zero-scale-factor universality at its upper critical dimension, with a specific form of logarithmic corrections to scaling (first proposed by Sachdev \textit{et al.} [Phys. Rev. B \textbf{50}, 258 (1994)]).
Our results conform to this expectation, but the logarithmic corrections to scaling do not match the form predicted by Sachdev \textit{et al.} 
We also show that the saturation transition becomes first order above a critical coupling ratio $(Q/J)_{\rm min}$ and is accompanied by magnetization jumps---metamagnetism. 
We obtain an exact solution for $(Q/J)_{\rm min}$ using a high magnetization expansion, and confirm the existence of the magnetization jumps beyond this value of coupling using quantum Monte Carlo simulations. 
\end{abstract}

\maketitle


\section{Introduction}

Models of quantum magnetism are of great interest in the quest to understand quantum phase transitions and many body states with strong quantum fluctuations. 
Studies in this field typically focus on identifying phases and phase transitions between them as a function of some coupling ratio. 
These coupling ratios are typically difficult or impossible to tune in experimental systems. 
In contrast, external magnetic fields are easy to adjust in experiments, making studies of field-driven quantum phase transitions particularly relevant. 
Despite this fact, such phase transitions have been largely neglected by the theoretical literature. 
Here, we present a study of the field-driven saturation transition in a two-dimensional (2D) quantum antiferromagnet known as the $J$-$Q$ model. 
In this model, a nearest neighbor antiferromagnetic Heisenberg exchange of strength $J$ competes with a four-spin interaction of strength $Q$, which favors valence-bond solid order. 
The form of this term is $-Q P_{i,j} P_{k,l}$ (where $P_{i,j} \equiv \frac{1}{4} - \mathbf{S}_i \cdot \mathbf{S}_j$ and $i,j$ and $k,l$ denote parallel bonds of an elementary plaquette of the square lattice). 
While the $Q$ interaction competes with the Heisenberg exchange ($-J P_{i,j}$), it does not produce frustration in the conventional sense, allowing numerically-exact quantum Monte Carlo studies of the physics. 
We find that the field-driven saturation transition from the antiferromagnet to the fully saturated state in the $J$-$Q$ model is composed of two regimes: a low-$Q$ continuous transition and high-$Q$ discontinuous (first order) transition with magnetization jumps, both of which will be address here. 

For low $Q$, we find that the transition is continuous and is therefore expected to be governed by a zero-scale-factor universality, which was predicted by Sachdev \textit{et al.} in 1994 \cite{sachdev1994}, but until now had not been tested numerically or experimentally in spatial dimension $d=2$ (2D). 
Although the leading order behavior matches the Sachdev \textit{et al.} prediction, we find multiplicative logarithmic violations of scaling at low temperature. Such violations are to be expected based on the fact that 2D represents the upper critical dimension for this transition, but these scaling violations do \textit{not} match the form predicted by Sachdev \textit{et al.} for reasons that are currently unclear.  

At high $Q$, the saturation transition is first order and there are discontinuities (jumps) in the magnetization known as \textit{metamagnetism} \cite{iaizzi2017,jacobs1967,stryjewski1977}. 
These jumps are caused by the onset of attractive interactions between magnons (spin flips on a fully polarized background) mediated by the $Q$-term (a mechanism previously established in the 1D $J$-$Q$ model \cite{iaizzi2017}). 
We use a high-magnetization expansion to obtain an exact solution for the critical coupling ratio $(Q/J)_{\rm min}$ where the jump first appears.

\section{Background}

The $J$-$Q$ model is part of a family of Marshall-positive Hamiltonians constructed from products of singlet projection operators \cite{kaul2013}. 
The two-dimensional realization of the \mbox{$J$-$Q$} model is given by 
\begin{eqnarray}
H_{JQ} & = & -J \sum \limits_{\braket{i,j}} P_{i,j} - Q \sum \limits_{\braket{i,j,k,l}} P_{i,j} P_{k,l} 
\end{eqnarray}
where $\braket{i,j}$ sums over nearest neighbors and $\braket{i,j,k,l}$ sums over plaquettes on a square lattice as pairs acting on rows $\begin{smallmatrix} k&l\\i&j\end{smallmatrix} $ and columns $\begin{smallmatrix} j&l\\i&k\end{smallmatrix}$ \cite{sandvik2007}. 
The zero-field $J$-$Q$ model has been extensively studied in both one \cite{iaizzi2017,sanyal2011,tang2011a,tang2014} and two \cite{sandvik2007,lou2009,sandvik2010,jin2013,tang2013} spatial dimensions, where it provides a numerically tractable way to study the deconfined quantum critical point marking the transition between the N\'{e}el antiferromagnetic state and the valence-bond solid (VBS). 
The VBS breaks $Z_4$ lattice symmetry to form an ordered arrangement of local singlet pairs. 
Here we will not focus on this aspect of the $J$-$Q$ model, but instead add an external magnetic field $h_z$,

\begin{align}
H_{JQh} = H_{JQ} - h_z \sum \limits_i S^z_i,
\end{align}
and study the magnetization near the field-driven transition to saturation. 
A separate paper \cite{iaizzi2018dqc} will discuss magnetic field effects in the vicinity of the N\'{e}el-VBS transition (see also Ref.~\onlinecite{mythesis}).  
Hereafter we will either fix the energy scale by (1) setting $J=1$ (and thereby referring to the dimensionless parameters $q\equiv Q/J$ and $h \equiv h_z/J$) or by (2) requiring $J+Q=1$ (and thereby referring to the dimensionless parameters $s\equiv Q/(J+Q)$ and $h\equiv h_z/(J+Q)$).  

The magnetization jumps correspond to a first-order phase transition (sometimes called metamagnetism) in which the magnetization changes suddenly in response to an infinitesimal change in the magnetic field \cite{jacobs1967,stryjewski1977}. 
This sort of transition usually occurs in spin systems with frustration or intrinsic anisotropy \cite{gerhardt1998,hirata1999,aligia2000,dmitriev2006,kecke2007,sudan2009,arlego2011,kolezhuk2012,huerga2014}, but recent work \cite{iaizzi2015,iaizzi2017,mythesis,mao2017} has shown that metamagnetism occurs in the 1D $J$-$Q$ model, which (in the absence of a field) is both isotropic and unfrustrated. 
The magnetization jumps in the 1D $J$-$Q$ model are caused by the onset of attractive interactions between magnons (flipped spins against a fully polarized background) mediated by the four-spin interaction \cite{iaizzi2017}. 
In the 1D case the critical coupling ratio $q_{\rm min}$ can be determined exactly using a high-magnetization expansion \cite{iaizzi2017}. 
Here we build on previous work \cite{iaizzi2017} to include the 2D case. 

Zero-scale-factor universality, first proposed by Sachdev \textit{et al.} in Ref.~\onlinecite{sachdev1994}, requires response functions to obey scaling forms that depend only on the \textit{bare} coupling constants, without any nonuniversal scale factors in the arguments of the scaling functions.
It applies to continuous quantum phase transitions that feature the onset of a nonzero ground state expectation value of a conserved density \cite{sachdev1994,iaizzi2017}. 
The saturation transition in the $J$-$Q$ model for $q<q_{\rm min}$ is just such a situation \cite{iaizzi2017}, although the 2D case is at the upper critical dimension of the theory, so we expect to find (universal) multiplicative logarithmic corrections to the zero-factor scaling form. 

\textit{Outline:} 
The methods used in this work are summarized in \cref{s:meta_methods}. 
In \cref{s:pd}, we discuss a schematic phase diagram of the 2D $J$-$Q$ model. 
In \cref{s:jump}, we focus on the onset of a magnetization jump at $q_{\rm min}$, where the saturation transition becomes first order, and derive an exact result for the value of $q_{\rm min}$.
In \cref{s:zsf2d} we discuss the universal scaling behavior near the continuous saturation transition, focusing on tests of the zero-scale-factor prediction as well as the presence of multiplicative logarithmic corrections expected at the upper critical dimension ($d=2$).
Our conclusions are discussed further in \cref{s:discussion}. 

\section{Methods \label{s:meta_methods}}

For the exact solution for $q_{\rm min}$ we have used Lanczos exact diagonalization \cite{sandvik2011computational} of the two-magnon (flipped spins on a fully polarized background) Hamiltonian, which we derive in an exact high-field expansion. 
The large-scale numerical results obtained here were generated using the stochastic series expansion quantum Monte Carlo (QMC) method with directed loop updates \cite{sandvik_dl} and quantum replica exchange.
This QMC program is based on the method used in our previous work \cite{iaizzi2017}. 
The stochastic series expansion is a QMC method which maps a \mbox{$d$-dimensional} quantum problem onto a $(d+1)$-dimensional classical problem by means of a Taylor expansion of the density matrix $\rho = e^{-\beta H}$, where the extra dimension roughly corresponds to imaginary time in a path-integral formulation \cite{sandvik2011computational}. 
In the QMC sampling, the emphasis is on the operators that move the world-lines rather than the lines themselves. 
The method used here is based on the techniques first described in Ref.~\onlinecite{sandvik_dl} (as applied to the Heisenberg Model). 

In addition to the standard updates, we incorporated quantum replica exchange \cite{hukushima1996,sengupta2002}, a multicanonical method in which the magnetic field (or some other parameter) is sampled stochastically by running many simulations in parallel with different magnetic fields and periodically allowing them to swap fields in a manner that obeys the detailed balance condition. 
To further enhance equilibration we used a technique known as $\beta$-doubling, a variation on simulated annealing. 
In $\beta$-doubling simulations begin at high temperature and the desired inverse temperature is approached by successive doubling of $\beta$; each time $\beta$ is doubled a new operator string is formed by appending the existing operator string to itself \cite{sandvik2002}. 
A detailed description of all of these techniques can be found in Chapter 5 of Ref.~\onlinecite{mythesis}.  

\section{Phase Diagram \label{s:pd}}

\begin{figure} 
\centering
\includegraphics[width=80mm]{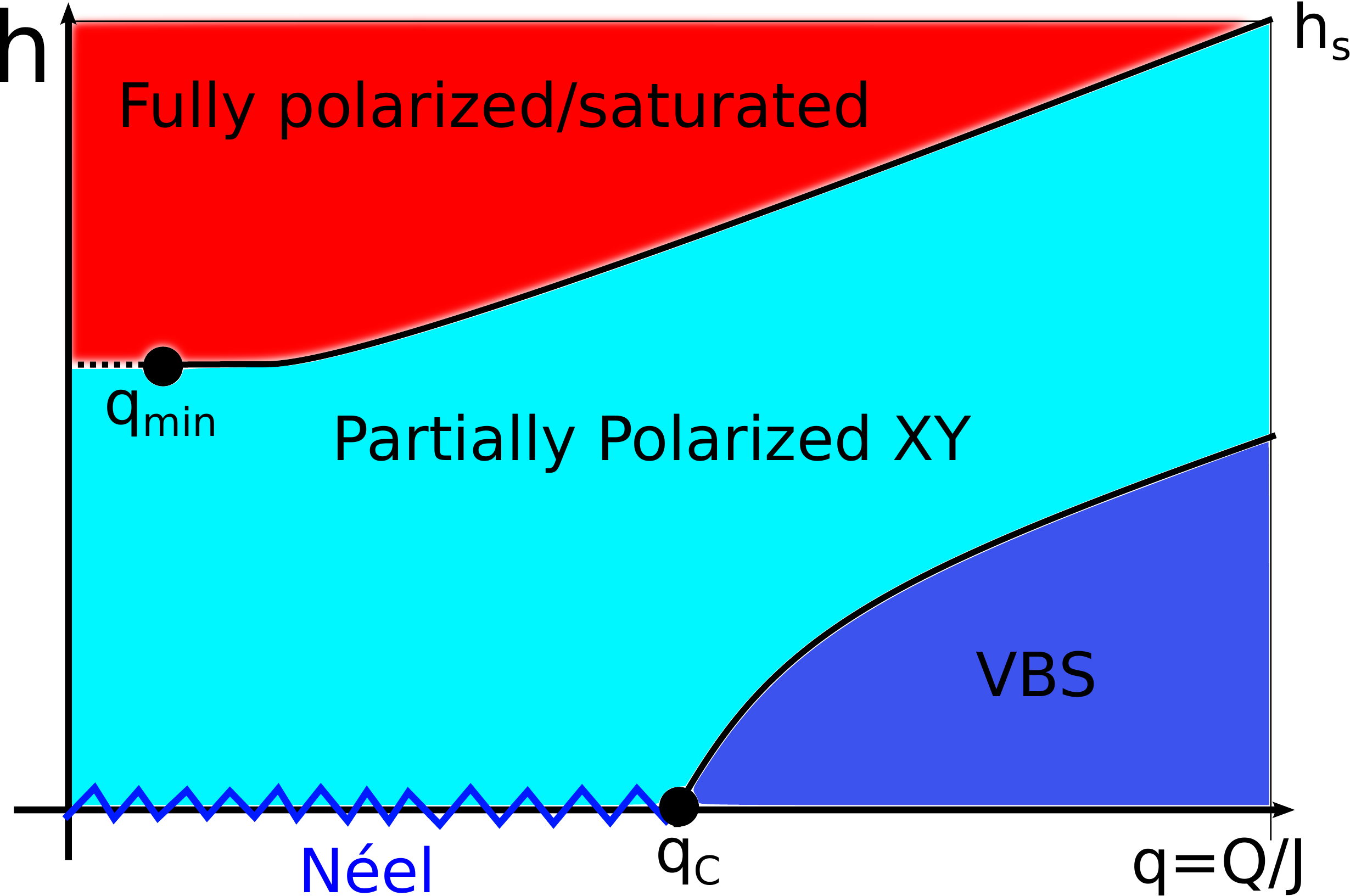}
\caption{Schematic phase diagram of the 2D \mbox{$J$-$Q$} model in an external field at zero temperature. The different phases and critical points are explained in the text. \label{f:phase}}
\end{figure}

In Fig.~\ref{f:phase}, we present a schematic zero-temperature phase diagram of the 2D \mbox{$J$-$Q$} model. 
Here the $h$ axis corresponds to the well-understood 2D Heisenberg antiferromagnet in an external field, and the $q$-axis corresponds to the previously studied \cite{sandvik2007,lou2009,sandvik2010,jin2013,tang2013,shao2016} zero-field \mbox{$J$-$Q$} model, which for $q<q_c$ has long-range antiferromagnetic N\'eel order in the ground state. 
At finite temperature $O(3)$ spin-rotation symmetry (which is continuous) cannot be spontaneously broken (according to the Mermin-Wagner Theorem \cite{mermin1966}), so there is no long-range spin order; instead there is a ``renormalized classical'' regime with the spin correlation length diverging exponentially as $T\rightarrow 0$ like $\xi \propto e^{2\pi \rho_s/T}$ \cite{chakravarty1988}.  
At $q_c$, the zero-field $J$-$Q$ model undergoes a quantum phase transition to the valence-bond solid (VBS) state. 
The off-axes area of \cref{f:phase} has not previously been studied; we here focus on the region around the field-driven saturation transition, $h_s(q)$.  
The region around the deconfined quantum critical point, $q_c$, will be addressed in a forthcoming publication \cite{iaizzi2018dqc}.  

Starting from the N\'{e}el state ($q<q_c$) on the $q$ axis, adding a magnetic field forces the antiferromagnetic correlations into the $XY$ plane, producing a partially polarized canted antiferromagnetic state. 
At finite temperature, there is no long-range N\'{e}el order, but the addition of a field permits a BKT-like transition to a phase with power-law spin correlations. 
For $q>q_c$, the ground state has VBS order. 
This state has a finite gap, so it survives at finite temperature and is destroyed by the magnetic field only after it the closes spin gap. 
The destruction of the VBS recovers the canted antiferromagnetic state (or partially polarized spin disordered phase for $T>0$).  

We here will focus on the saturation transition in the high-field region of the phase diagram. 
The system reaches saturation (where all spins are uniformly aligned in the $+z$ direction) at $h=h_s(q)$. 
For $q<q_{\rm min}$, this transition is continuous and the saturation field is given by $h_s(q\leq q_{\rm min})=4J$ (in this regime $h_s(q)$ is a dashed line). 
For $q>q_{\rm min}$ the saturation transition is first order (i.e. metamagnetic) and there are macroscopic discontinuities in the magnetization (in this regime $h_s(q)$ is a solid line). 
The point $q_{\rm min}$ denotes the onset of metamagnetism, here the magnetization is still continuous, but the magnetic susceptibility diverges at saturation (corresponding to an infinite-order phase transition). 
 
\section{Metamagnetism \label{s:jump}}

Magnetization jumps (also known as metamagnetism) can appear due to a variety of mechanisms including broken lattice symmetries, magnetization plateaus \cite{honecker2004}, localization of magnetic excitations \cite{richter2004,schnack2001,schulenburg2002}, and bound states of magnons \cite{aligia2000,kecke2007,iaizzi2017}. 
It has previously been established that magnetization jumps occur in the \mbox{$J$-$Q$} chain caused by the onset of a bound state of magnons \cite{iaizzi2015,iaizzi2017,mao2017}; to our knowledge, this is the first known example of metamagnetism in the absence of frustration or intrinsic anisotropy. 
To understand the mechanism for metamagnetism, we consider bosonic spin flips (magnons) on a fully polarized background. 
These magnons are hardcore bosons that interact with a short-range repulsive interaction in the Heisenberg limit. 
The introduction of the $Q$-term produces an effective short-range \textit{attractive} interaction between magnons. 
At $q_{\rm min}$, this attractive force dominates and causes pairs of magnons to form bound states. 

\subsection{Exact Solution for $q_{\rm min}$ \label{s:jumpED}}

We will now find $q_{\rm min}$ for the 2D $J$-$Q$ model using the procedure developed for the $J$-$Q$ chain in Ref.~\onlinecite{iaizzi2017}. 
Let us define bare energy of an $n$-magnon state, $\bar E_n$, as

\begin{align}
E_n (J,Q,h) = \bar E_n (J,Q) -nh/2.
\end{align}
We can then define the binding energy of two magnons as

\begin{align}
\Xi(q) \equiv 2 \bar E_1 - \bar E_2. 
\end{align}
The $Q$ term is nonzero only when acting on states where there are exactly two magnons on a plaquette, so it does not contribute to the single-magnon dispersion, which has a tight-binding-like form \cite{iaizzi2017}. 
We can therefore solve analytically for the single-magnon energy, $\bar E_1 = -4J$. 
The two-magnon energy, $\bar E_2$ corresponds to the ground state in the two magnon sector, and must be determined numerically. 
Since this is only a two-body problem, relatively large systems can be studied using Lanczos exact diagonalization to obtain $\bar E_2$ to arbitrary numerical precision. 

\begin{figure}
\centering
\includegraphics[width=80mm]{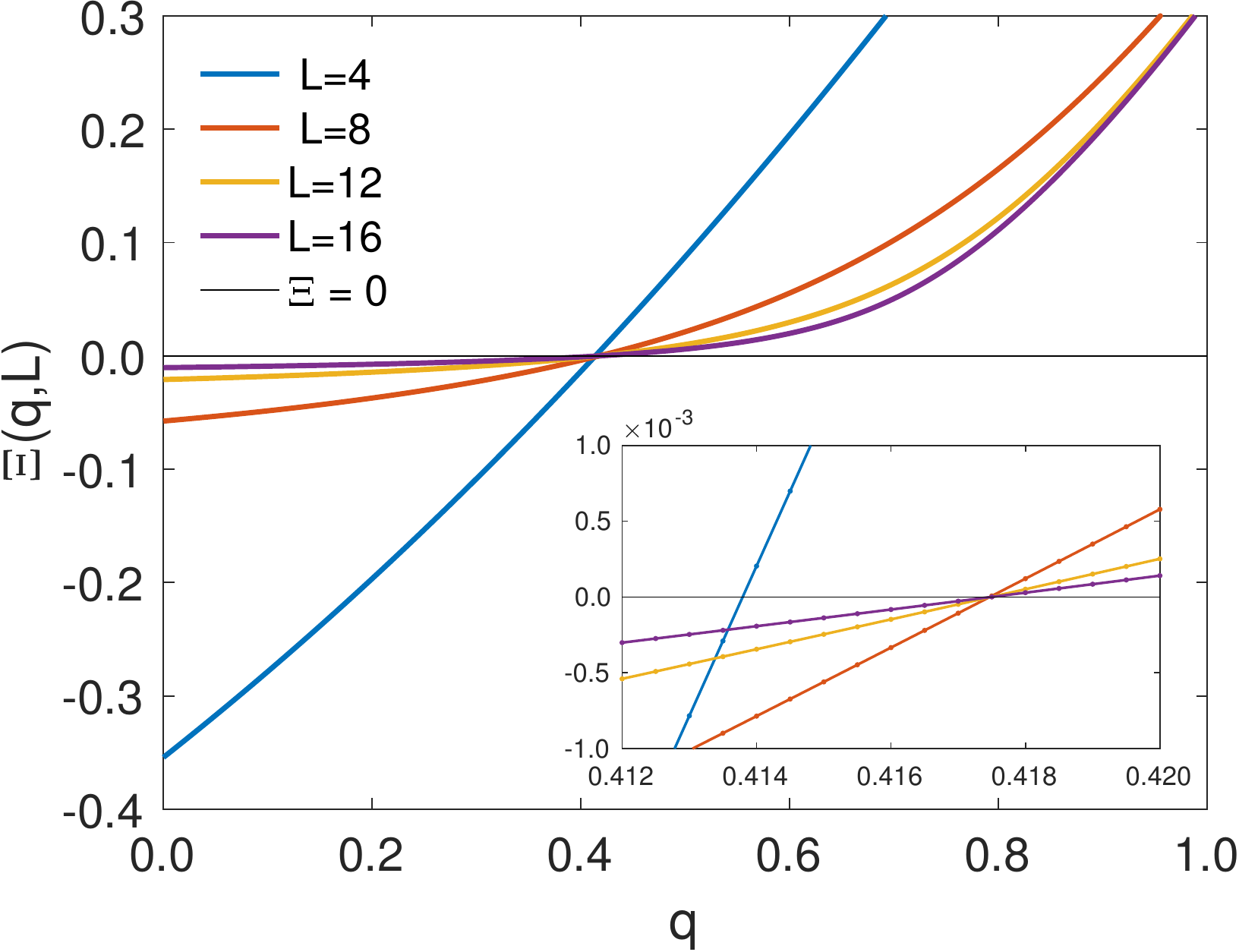}
\caption{Binding energy $\Xi(q,L)$ plotted against $q$ for several system sizes calculated using exact diagonalization. The thin black line represents $\Xi=0$. Inset: zoomed-in view of crossing point. \label{f:xing}}
\end{figure}

In Fig.~\ref{f:xing} we plot the binding energy of two magnons, $\Xi(q,L)$, for $0 \leq q \leq 1$ and $L=4, \, 8, \, 12, \, 16$. 
For all sizes the binding energy becomes positive around $q\approx 0.417$. 
We can also see that Fig.~\ref{f:xing} strongly resembles the analogous figure for the $J$-$Q$ chain (see Fig.~6 of Ref.~\onlinecite{iaizzi2017}).
For $q<q_{\rm min}$ finite size effects result in an \textit{underestimate} of the binding energy and for $q>q_{\rm min}$ finite size effects cause an \textit{overestimate} of the binding energy. 
Around $q_{\rm min}$ these effects cancel out and the crossing is \textit{nearly} independent of system size (in the 1D case the crossing is exactly independent of $L$). 
Using a bracketing procedure, we can extract $q_{\rm min}(L)$ to arbitrary numerical precision. 
\cref{t:qmin} contains a list of $q_{\rm min}(L)$ for select $L \times L$ systems with $L\leq 24$. 
$q_{\rm min}$ converges exponentially fast in $L$, so based on extrapolation using these modest sizes we know $q_{\rm min}(L=\infty) = 0.41748329$ to eight digits of precision. 
Although we do not plot it here, the exponential convergence of $q_{\rm min}(L)$ can be seen from the underlines in \cref{t:qmin}, which indicate the digits which are converged to the thermodynamic limit; the number of underlined digits grows linearly with $L$. 
Note here that $q_{\rm min}$ is not the same as $q_c$ (the N\'{e}el-VBS transition point), and these two phase transitions are governed by completely different physics. 

\begin{table}
\caption{$q_{\rm min}(L)$ calculated to machine precision for select $L \times L$ systems using Lanczos exact diagonalization. The underlined portions of the numbers represent the digits that are fully converged to the thermodynamic limit. \label{t:qmin}}
\begin{center}
\begin{tabular}{c|c}
L 	&	$q_{\rm min}$ \\
\hline
 4	&	\underline{0.41}3793103448 \\
 6	&	\underline{0.417}287630402 \\
 8	&	\underline{0.4174}67568061 \\
10	&	\underline{0.41748}1179858 \\
12	&	\underline{0.417482}857341 \\
14	&	\underline{0.417483}171909 \\
16	&	\underline{0.4174832}50752 \\
18	&	\underline{0.4174832}74856 \\
20	&	\underline{0.4174832}83375 \\
22	&	\underline{0.41748328}6742 \\
24	&	\underline{0.41748328}8198 \\
\end{tabular}
\end{center}
\end{table}

\begin{figure}
\centering
\includegraphics[width=80mm]{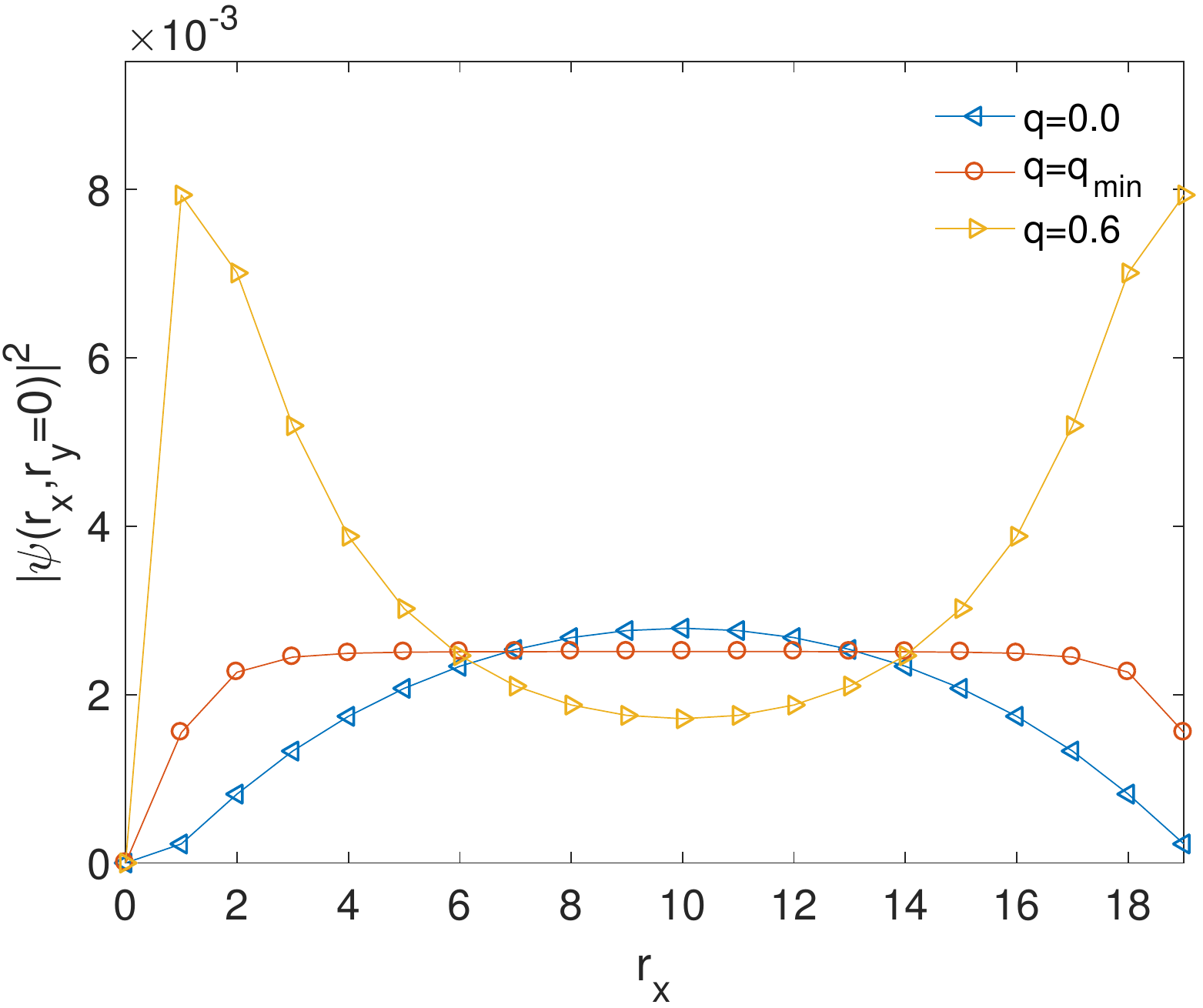}
\caption{Probability density of magnon separation in the $x$-direction for $r_y = 0$, $|\psi(r_x,r_y=0)|$ in the two-magnon sector of the $J$-$Q$ model; calculated using Lanczos exact diagonalization. \label{f:pdens}}
\end{figure}

In \cref{f:pdens} we plot the ground state probability density in the two magnon sector as a function of separation of the magnons in the $x$ direction, $r_x$ (with $r_y =0$).  
Here we consider a small ($18\times18$) system to make the features at the boundary easier to distinguish on the scale of the figure. 
For $q=0$, we can see that the probability density takes on the form of a free particle with periodic boundary conditions in $r_x,r_y$, with a single excluded site at $r_x=r_y=0$. 
In the continuum limit, this corresponds to a repulsive delta potential. 
For $q>q_{\rm min}$ the wave function takes on the exponentially decaying form of a bound state. 
At $q=q_{\rm min}$ (the crossover between repulsive and attractive interactions) the wave function becomes flat with an exponentially-decaying short-distance disturbance of the form $\psi \propto 1-a e^{-r_x /b}$ (this was confirmed by further data not depicted here).
This exponential disturbance explains why the finite size effects vanish exponentially near $q_{\rm min}$. 
This form of the  wave function in the 2D case stands in contrast to the flat wave function in the 1D $J$-$Q$ model, where the bulk wavefunction at $q_{\rm min}$ is perfectly flat and $q_{\rm min}$ is exactly independent of $L$ for $L>6$ \cite{iaizzi2017}. 

The onset of attractive interactions between magnons has previously been found to cause metamagnetism \cite{aligia2000,kecke2007,iaizzi2017}, but bound pairs of magnons are not a sufficient condition to guarantee the existence of a macroscopic magnetization jump. 
The magnetization could, for example, change by steps of $\Delta m_z=2$, but never achieve a macroscopic jump \cite{honecker2006,kecke2007}. 
For a true jump to occur, the point $q_{\rm min}$ must be the beginning of an instability leading to ever larger bound states of magnons. 
In the next section we will confirm numerically that a macroscopic magnetization jump does in fact occur in the full magnetization curves obtained via quantum Monte Carlo simulations. 
It will not be possible to detect the onset of the magnetization jump (which is initially infinitesimal) by directly examining the magnetization curves due to finite-temperature rounding. 
Instead, in \cref{s:zsf2d} we will examine the scaling of the magnetization near saturation and find that a qualitative change in behavior consistent with the onset of a different universality class, occurs at the predicted value of $q_{\rm min}$. 

\subsection{Quantum Monte Carlo Results}

\begin{figure}
\centering
\includegraphics[width=80mm]{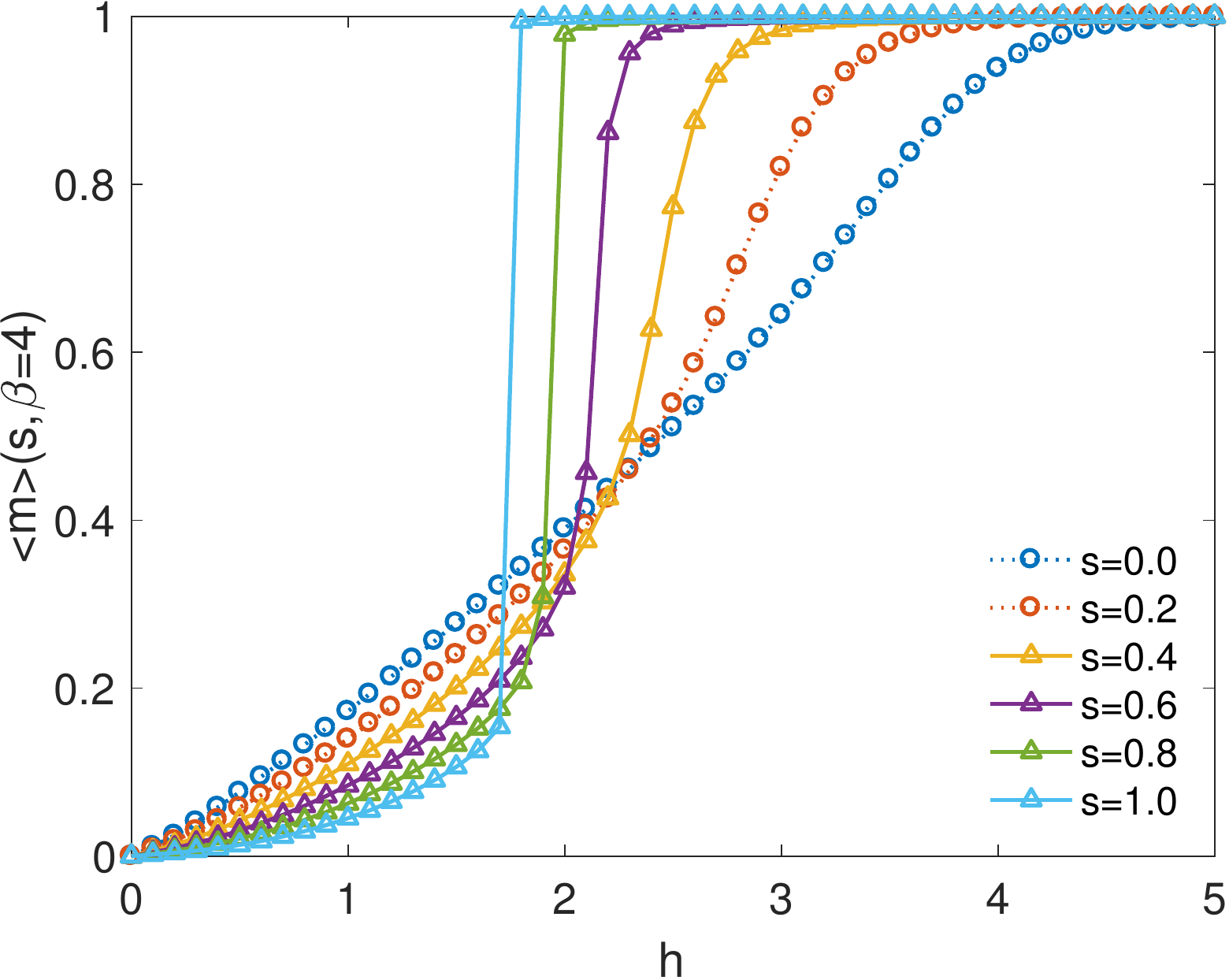}
\caption{Magnetization density of the 2D \mbox{$J$-$Q$} model as function of external field, $h$, for a range of different values of $s$ defined such that $J=1-s$ and $Q=s$. Here $s=0,0.2,0.4,0.6,0.8,1$ with $\beta=4$ correspond to $q=0, 0.25, 0.67, 1.5, 4, \infty$, respectively (with rescaled non-constant $\beta$). Results from QMC with quantum replica exchange. \label{f:mmag}}
\end{figure}

In Fig.~\ref{f:mmag}, we plot the magnetization density,

\begin{align}
m = \frac{2}{L^2}\sum S^z_i,
\end{align}
of the 2D \mbox{$J$-$Q$} model as a function of external field for several different values of \mbox{$0\leq s \leq 1$} where $s$ is defined such that \mbox{$J=1-s$} and $Q=s$ such that \mbox{$J+Q=1$}. 
Here we use a $16\times 16$ lattice with $\beta=4$. 
Ordinarily, QMC can study much larger systems than this, but as was observed in our previous work \cite{iaizzi2015,iaizzi2017}, the $J$-$Q$ model with a field is exceptionally difficult to study, even when using enhancements such as $\beta$-doubling and quantum replica exchange (both used here). 
We have compared to smaller and larger sizes and finite size effects do not qualitatively affect the results on the scale of Fig.~\ref{f:mmag}. 
For $s=0$ (the Heisenberg limit), the magnetization is linear in $h$ for small fields, and smoothly approaches saturation at $h=4J$. 
When $s=0.2$, corresponding to a coupling ratio of $q=0.25$, the magnetization curve begins to take on a different shape: shallower at low field and steeper near saturation. 
This trend continues as $s$ increases: for $s\geq 0.8$, there is a clear discontinuity. 
Although the jump should appear for $q \geq q_{\rm min} = 0.417$, which corresponds to $s_{\rm min} = 0.294$, this is difficult to distinguish in the QMC data. 
At $q_{\rm min}$, the jump is infinitesimal, and even when the jump is larger, such as for $s=0.4$ and $s=0.6$, it is hard to clearly distinguish due to finite temperature effects, which round off the discontinuity in the magnetization. 
These results are nonetheless consistent with the value of $q_{\rm min}$ predicted using the exact method, and demonstrate that a macroscopic magnetization jump does in fact occur. 
We will discuss more evidence for $q_{\rm min} \approx 0.417$ from the critical scaling of the magnetization near saturation in \cref{s:zsf2d}. 

\section{Zero-Scale-Factor Universality\label{s:zsf2d}}

In the $J$-$Q$ model, magnetization near saturation should be governed by a remarkably simple zero-scale-factor universality for $q<q_{\rm min}$ (where the saturation transition is continuous) \cite{sachdev1994,iaizzi2017}. 
Here, ``zero-scale-factor'' means that the response functions are universal functions of the bare coupling constants and do not depend on any nonuniversal numbers \cite{sachdev1994}. 
Zero-scale-factor universality applies to low-dimensional systems where there is a quantum phase transition characterized by a smooth onset of a conserved density \cite{sachdev1994}. 
Typically this is applied to the transition from the gapped singlet state of integer spin chains to a field-induced Bose-Einstein condensate of magnons (excitations above the zero magnetization state). 
In the $J$-$Q$ model, we instead start from the saturated state with $h>h_s$, and consider flipped spins on this background---magnons---as $h$ is decreased below $h_s$. 
In the 1D case, the zero-factor scaling form applies for all $q<q_{\rm min}$ at sufficiently low temperature, and is violated by a logarithmic divergence at exactly $q_{\rm min}$  \cite{iaizzi2017}. 
The 2D $J$-$Q$ model is at the upper critical dimension of this universality class, so we expect multiplicative logarithmic violations of the zero-factor scaling form for all $q$.  
We will describe the universal scaling form and its application to the saturation transition in the 2D $J$-$Q$ model and then show that the low-temperature violations of the scaling form do not match the prediction in Ref. \onlinecite{sachdev1994}. 

In two spatial dimensions, the zero-factor scaling form for the deviations from saturation $(\delta \braket{m} \equiv 1 - \braket{m})$ is given by Eq. (1.23) of Ref. \onlinecite{sachdev1994}:

\begin{align}
\delta \braket{m} = g \mu_B \left( \frac{2M}{\hbar^2 \beta}\right) \mathcal{M}(\beta \mu), \label{sachdev}
\end{align}
where $M$ is the bare magnon mass (which is $M=1$ when $J=1$), and $\mu$ represents the field, $\mu \equiv h_s-h$. 
For $q\leq q_{\rm min}$, the saturation field is $h_s=4J$ (which can be determined analytically from the level crossing between the saturated state and the state with a single flipped spin \cite{iaizzi2017}). 
We set $\hbar=1$ and $\delta \braket{m} = g \mu_B \braket{n}$ to define the rescaled magnon density:

\begin{align}
n_s (q,\beta \mu) \equiv \frac{\beta \braket{n}}{2} = \mathcal{M}(\beta \mu). \label{ns}
\end{align}
We emphasize again that these magnons are spin flips on fully polarized background, so $n\rightarrow 0$ corresponds to the saturated state. 
The field is also reversed from the usual case (of a gapped singlet state being driven to a polarized ground state by applying a uniform field). 
Thus, in the present case, $h>h_s$ produces a negative $\mu$, which means $n\rightarrow0$, and $h<h_s$ corresponds to a positive $\mu$ and a finite density of magnons. 

At the saturation field, $\mu=0$, the scaling form in \cref{ns} predicts that the density takes on a simple form:

\begin{align}
\braket{n} = 2 \mathcal{M}(0) T.
\end{align}
At this same point the rescaled density, $n_s$, becomes independent of temperature:

\begin{align}
n_s (q,0) \equiv \frac{\beta \braket{n}}{2} = \mathcal{M}(0).
\end{align}
In our case there are two spatial dimensions and $z=2$ imaginary time dimensions, so the total dimensionality is $d=4$, which is the upper critical dimension of the zero-scale-factor universality \cite{sachdev1994}. 
At low temperatures, we therefore expect to see multiplicative logarithmic violations of this scaling form, which should be universal as well. 

\begin{figure}
\centering
\includegraphics[width=80mm]{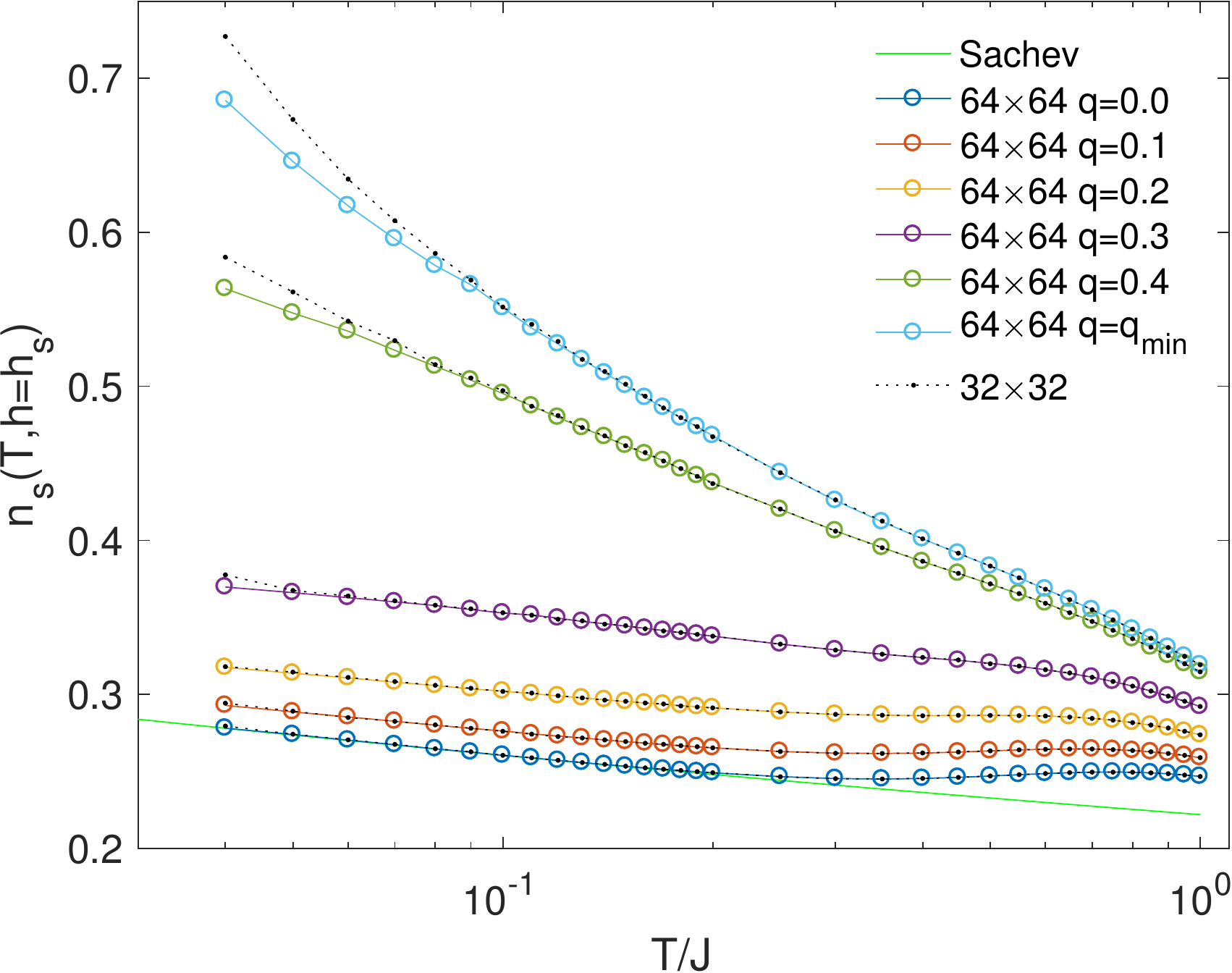}
\includegraphics[width=80mm]{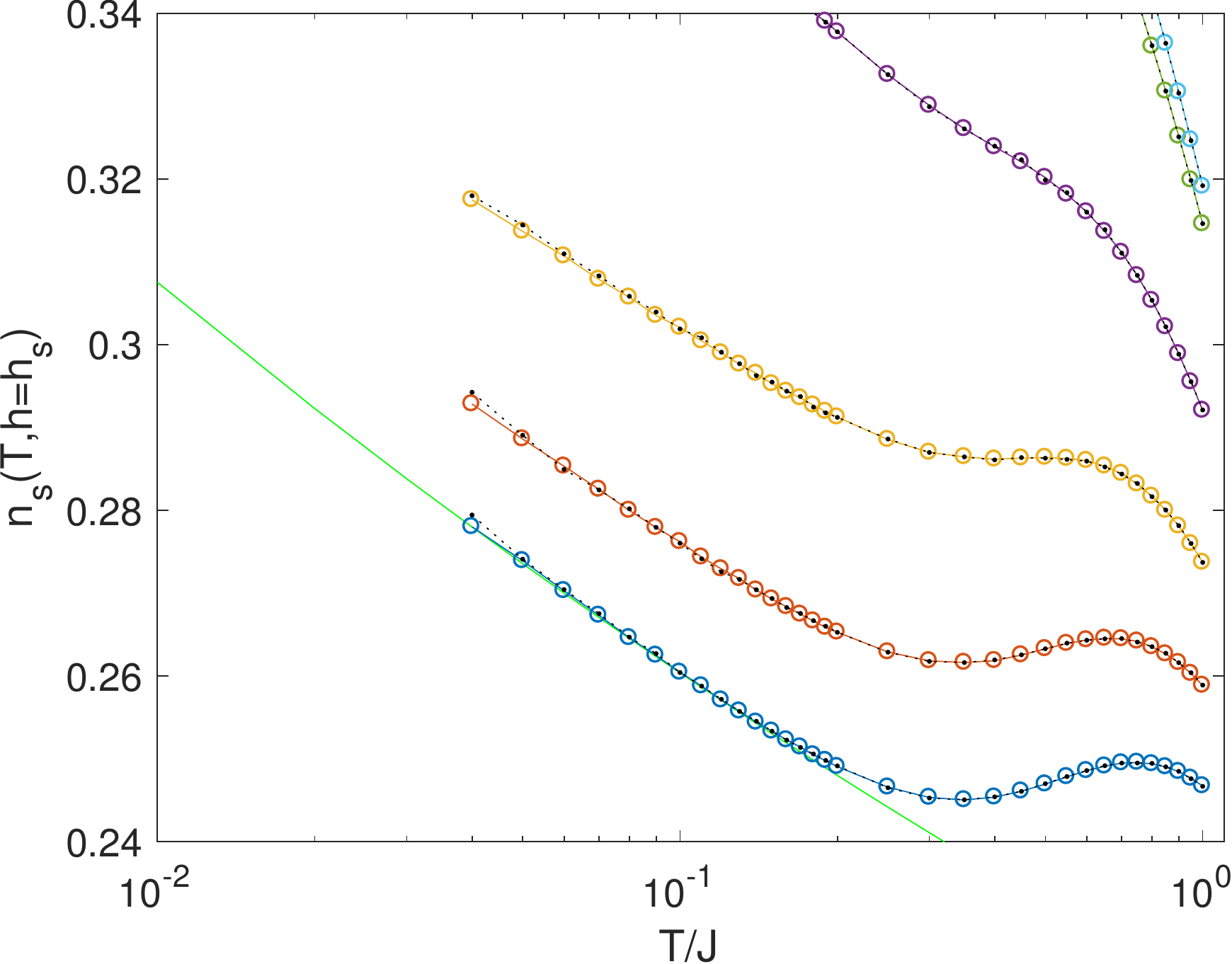}
\caption{(Top) The zero-scale-factor-rescaled magnon density [Eq.~(\ref{ns})] at $h=h_s$, $\mu=0$ calculated using QMC with quantum replica exchange. The bright green line is a fit to the log-corrected scaling form Eq.~(\ref{logfit}). (Bottom) A zoomed-in view. \label{f:zsf}}
\end{figure}

In Fig. \ref{f:zsf}, we plot the rescaled magnon density at saturation, $n_s(q,\mu=0)$, as a function of temperature for two different sizes, $32\times32$ and $64\times64$. 
Here we use the exact value of the saturation field $h_s(q\leq q_{\rm min}) = 4J$. 
These sizes are large enough that finite size effects only become important at low temperature; the results for the two different sizes overlap completely for $T\geq 0.1$, but exhibit some separation at lower temperature depending on the value of $q$. 
From simulations of $96 \times 96$ and $128 \times 128$ systems (not depicted here) we know that the $64 \times 64$ curve for $q=q_{\rm min}$ is converged to the thermodynamic limit within error bars.

If there were no corrections to \cref{ns} the lines in \cref{f:zsf} would exhibit no temperature dependence. 
Instead, we observe violations of the scaling form for all $q$. 
For $q=0$, there is some non-monotonic behavior, with a local minimum around $T=0.35$; at low temperatures, $n_s(T)$ appears to diverge like $\log(1/T)$, which on this semi-log scale manifests as a straight line. 
For $q=0.1$ and $0.2$, the behavior is similar, although the whole curve is shifted upwards. 
For $q=0.3$, the local minimum in $n_s(T)$ appears to be gone. 
The divergence for $q<q_{\rm min}$ looks log-linear, but it is difficult to distinguish between different powers of the log by fitting alone. 
At $q=0.4$ and $q=q_{\rm min}=0.4174833$, finite size effects become more important, and it appears that the log has a different power. 

\subsection{Behavior around $q_{\rm min}$}

We can also use the low-temperature behavior of $n_s$ in \cref{f:zsf} to verify our prediction of $q_{\rm min}$ (from the high-magnetization expansion discussed in \cref{s:jumpED}). 
At $q_{\rm min}$, the transition is no longer the smooth onset of a conserved density, therefore the zero-scale-factor universality does not apply (not even with logarithmic corrections).  
For all $q<q_{\rm min}$, the low-temperature divergence appears to obey a form $\log \left( \frac{1}{T} \right)$, or some power of it. 
At $q=q_{\rm min}$ the divergence of $n_s(q_{\rm min},T)$ takes on a \textit{qualitatively} different form that appears to diverge faster than $\log \left( \frac{1}{T} \right)$. 
This confirms the value of $q_{\rm min}$ predicted by the high-magnetization expansion, even though no sign of a discontinuity can be observed in the magnetization curves themselves due to finite-temperature rounding (see \cref{f:mmag}). 

\subsection{Low-temperature scaling violations}

Sachdev \textit{et al.} \cite{sachdev1994} derived a form for the logarithmic violations of the zero-scale-factor universality that occur at the upper critical dimension. 
At $\mu =0$ (saturation, $h=h_s$) they predict that the magnon density will take on the form

\begin{align}
\braket{n} = \frac{2M k_B T}{4 \pi} \left[ \log \left( \frac{\Lambda^2}{2M k_B T} \right) \right]^{-4}
\end{align}
(see Eq. (2.20) of Ref.~\onlinecite{sachdev1994}). 
Where $\Lambda$ is an upper (UV) momentum cutoff. 
We can plug this into Eq.~(\ref{ns}) to find a prediction for the log-corrected form of the rescaled magnon density:

\begin{align}
n_s(\mu =0) = \frac{M}{4 \pi} \left[ \log \left( \frac{\Lambda^2}{2M k_B T} \right) \right]^{-4} .\label{eq:logcorr}
\end{align}
This form should also be universal, but the UV cutoff should depend on microscopic details. 

For simplicity we will restrict our analysis to the Heisenberg limit ($Q=0$). 
Setting the magnon mass, $M=1$ (the bare value) and introducing a fitting parameter, $a$, we can attempt to fit our QMC results for $n_s(q=0,T\rightarrow0)$ to the form 

\begin{align}
n_s = a \left[ \log \left( \frac{\Lambda^2}{T} \right) \right]^{-4} \label{logfit}.
\end{align}
Automatic fitting programs were unable to find suitable values of $a$ and $\Lambda$ (in the low temperature regime where the divergence appears), so we manually solved for $a$ and $\Lambda$ using two points: $n_s(T=0.04)=0.278$ and $n_s(T=0.10)=0.2604$, finding $a=2.65354 \times 10^6$ and $\Lambda = 1.7\times10^{-13}$. 
We plot the resulting curve as a bright green line in Fig.~\ref{f:zsf}. 
Although this \textit{appears} to produce a good fit to the rescaled numerical data at low $T$, the fitting parameters do not make physical sense. 
The prefactor is fixed by the theory to be $a=M/(4 \pi)\approx 0.08$, yet the fitted value is huge: $a\approx 10^6$ (7 orders of magnitude too large). 
Worse yet, the UV cutoff, $\Lambda$, is extremely small ($10^{-13}$), much smaller than any other scale in this problem. 
In zero-scale-factor universality, there should be no renormalization of bare parameters, but even allowing for renormalization of the mass, $M$ (perhaps due to being at the upper critical dimension), it is not possible for \cref{logfit} to match the data while maintaining a physically sensible (i.e., large) value of the UV cutoff $\Lambda$. 

On close inspection, the fit in Fig.~\ref{f:zsf} bears a remarkable resemblance to a linear $\log T$ divergence. 
Indeed, since $T \gg \Lambda^2$, we can expand Eq.~(\ref{logfit}) in a Taylor series around small $u=\log T$ and we find an expression,
\begin{small}%
\begin{align}%
n_s = \frac{a}{\left( \log \Lambda^2 \right)^4} \left[ 1 + 4 \frac{\log T}{\log \Lambda^2} + 10 \left( \frac{\log T}{\log \Lambda^2} \right)^2 + \cdots \right],
\end{align}%
\end{small}%
that is linear in $\log T$ to first order and converges rapidly because $\log \Lambda^2 \approx -58$. 
Considering this fact and the unphysical parameters required to make the Sachdev form fit the data, it is clear that \cref{eq:logcorr} does not accurately describe the violations of the zero-scale-factor universality at its upper critical dimension. 
The apparent fit is instead a roundabout approximation of the true form, which is (approximately) proportional to $\log \left( \frac{1}{T} \right)$ to some positive power close to 1, although the exact power is difficult to determine from fitting. 
The reasons for the failure of the form predicted in Ref.~\onlinecite{sachdev1994} are unclear at this time. 

\section{Conclusions \label{s:discussion}}

Here we have presented a numerical study of the two-dimensional $J$-$Q$ model in the presence of an external magnetic field, focusing on the field-induced transition to the saturated (fully polarized) state. 
Building on a previous version of this study which focused on the 1D case \cite{iaizzi2015,iaizzi2017}, we have found that the saturation transition is metamagnetic (i.e., has magnetization jumps) above a critical coupling ratio $q_{\rm min}$. 
The existence of metamagnetism in the $J$-$Q$ model is surprising because all previously known examples of metamagnetic systems had either frustration or intrinsic anisotropy. 
This transition is caused by the onset of bound states of magnons (flipped spins against a fully polarized background) induced by the four-spin $Q$ term. 
The same mechanism can explain presence of metamagnetism in a similar ring-exchange model \cite{huerga2014}. 
We have determined $q_{\rm min}$ using an exact high-magnetization expansion (see Ref.~\onlinecite{iaizzi2017}). 
Although it is not possible to directly observe the onset of the magnetization jump in the QMC data, we do see an apparent change in universal scaling behavior at $q_{\rm min}$ (\cref{f:zsf}) which most likely corresponds to the presence of an infinitesimal magnetization jump which goes on to become the macroscopic jump we see at high $q$ and matches the results of our exact calculation. 
We cannot exclude the possibility that there is some intermediate behavior, like a spin nematic phase \cite{orlova2017} between $q\approx 0.417$ and some higher-$q$ onset of metamagnetism, but we believe this is unlikely. 

For $q<q_{\rm min}$, the saturation transition is continuous and is governed by a zero-scale-factor universality at its upper critical dimension \cite{sachdev1994}. 
This universality has already been shown apply to the 1D case \cite{iaizzi2017}. 
We have presented the first-ever numerical test of the zero-scale-factor universality in two dimensions. 
We found that the low-temperature scaling violations do \textit{not} obey the form proposed by Ref. \onlinecite{sachdev1994}, which predicts a divergence as a \textit{negative} power of $\log T$ as $T\rightarrow0$, and instead they appear to diverge as some \textit{positive} power of $\log T$. 

There are still some important unanswered questions here that need to be addressed in future studies. 
It is still unclear why the scaling violations to not match the form predicted by Ref.~\onlinecite{sachdev1994} or what should be the correct form of the violations. 
In a preliminary report (Ref. \onlinecite{mythesis}, Ch. 3) we considered an alternative form of the violations based on an analogy to the scaling of the order parameter in the 4D Ising universality class (also at its upper critical dimension). 
This universality matches the leading-order scaling predictions of the zero-scale-factor universality, and produced a better but not fully convincing agreement with the scaling violations observed in our QMC results. 
The theoretical basis for the analogy was weak. 
Further theoretical work is required to determine the correct form of the scaling violations. 
Once the proper form of the scaling violations is established it should be checked over the full range of its validity $0\leq q < q_{\rm min}$. 
At $q_{\rm min}$, the zero-scale-factor universality does not apply, but it is not currently clear what universal behavior should appear. 
Finally, we have not discussed the behavior of this system at low fields; this aspect of the $J$-$Q$ model including the field effect near the deconfined quantum critical point $q_c$ \cite{shao2016} will be addressed in a forthcoming publication \cite{iaizzi2018dqc}. 

\section*{Acknowledgements}
The work of A.I. and A.W.S. was supported by the NSF under Grants No. DMR-1710170 and No. DMR-1410126, and by a Simons Investigator Award.
A.I. acknowledges support from the APS-IUSSTF Physics PhD Student Visitation Program for a visit to K.D. at the Tata Institute of Fundamental Research in Mumbai. 
The computational work reported in this paper was performed on the Shared Computing Cluster administered by Boston University's Research Computing Services.  

\bibliography{bibstuff}

\end{document}